\documentclass[aps,prb,twocolumn,superscriptaddress,floatfix]{revtex4}

\usepackage{amsmath,amssymb,bm}
\usepackage{graphicx}

\begin{document}

\title{Half-Metallic Graphene Nanoribbons}
\author{Young-Woo Son} 
\affiliation
{Department of Physics, 
University of California at Berkeley, 
Berkeley, California 94720, USA}
\affiliation
{Materials Sciences Division, 
Lawrence Berkeley National Laboratory, Berkeley, 
California 94720, USA}
\author{Marvin L. Cohen} 
\affiliation
{Department of Physics, 
University of California at Berkeley, 
Berkeley, California 94720, USA}
\affiliation
{Materials Sciences Division, 
Lawrence Berkeley National Laboratory, Berkeley, 
California 94720, USA}
\author{Steven G. Louie}
\affiliation
{Department of Physics, 
University of California at Berkeley, 
Berkeley, California 94720, USA}
\affiliation
{Materials Sciences Division, 
Lawrence Berkeley National Laboratory, Berkeley, 
California 94720, USA}
\date{{\it submitted}, March 25, 2006}
\maketitle
{\bf Electrical current can be completely spin polarized in a class of materials known as 
half-metals, as a result of the coexistence of metallic nature for electrons 
with one spin orientation
and insulating for electrons with the other. 
Such asymmetric electronic states for the 
different spins have been predicted for some ferromagnetic metals$-$for example, the 
Heusler compounds~\cite{1}$-$and were first observed in a manganese perovskite~\cite{2}. 
In view of the potential for use of this property in realizing spin-based 
electronics, substantial efforts have been made to search for 
half-metallic materials~\cite{3,4}. 
However, organic materials have hardly been investigated in this context 
even though carbon-based nanostructures hold significant promise for future 
electronic device~\cite{5}. Here we predict half-metallicity in nanometre-scale 
graphene ribbons by using first-principles calculations. 
We show that this phenomenon is realizable 
if in-plane homogeneous electric fields are applied across the zigzag-shaped 
edges of the graphene nanoribbons, and that their magnetic property can be 
controlled by the external electric fields. The results are not only of 
scientific interests in the interplay between electric fields and electronic spin degree of 
freedom in solids~\cite{6,7} 
but may also open a new path to explore spintronics~\cite{3} at nanometre scale,
based on graphene~\cite{8,9,10,11}
}

When a single graphite layer is terminated by 
zigzag edges on both sides, which we refer here to as 
a zigzag graphene nanoribbon (ZGNR) (Fig. 1), 
there are peculiar localized electronic states at each edge~\cite{12,13}. 
These edge states (which are extended along the edge direction)
decay exponentially into the center of the ribbon, 
with decay rates depending on their momentum~\cite{12,13,14,15}. 
Such states have been observed in monoatomic step edges 
of graphite by using scanning probe techniques~\cite{16,17}. 
The localized edge states form a two-fold degenerate flat band 
at the Fermi energy ($E_F$), existing in about one third of 
the Brillouin zone away from the zone center~\cite{12,13,14,15}. 
By invoking band ferromagnetism, it has been suggested 
that an opposite spin orientation across the ribbon 
between ferromagnetically ordered localized edge states on 
each edge in ZGNRs is the ground-state spin configuration; 
that is, the total spin is zero~\cite{12,18,19}.  
Because the states around $E_F$ are the edge states 
and linear combinations of them, the effects of external transverse
fields are expected to be significant on these states, in constrast with
those on the extended states~\cite{20}.

\begin{figure}[b]
\centering
\includegraphics[width=8.5cm]{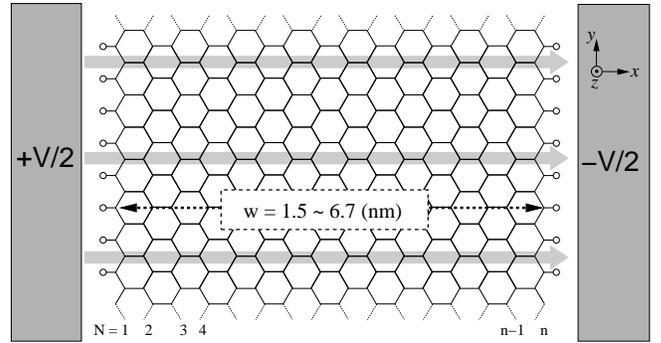}
\caption{
{\bf Graphene nanoribbon in electric fields.} 
Diagram of a zigzag 
graphene nanoribbon (ZGNR) with external transverse electric field
$E_\text{ext}$. $E_\text{ext}$ is 
applied across the ZGNR along the lateral direction ($x$ direction) 
in an open circuit split-gate configuration 
and is positive towards the right side. The ZGNRs 
are assumed to be infinite along the $y$ direction. 
A small longitudinal source-drain 
field could be applied to generate spin-polarized currents along the $y$-direction. 
Hydrogen atoms on the edges are denoted by circles. The ZGNR shown in this 
figure is an 16-ZGNR ($n$=16). The width $w$ of ZGNRs under study 
was in the range 1.5 $\sim$ 6.7 nm. 
}
\end{figure}

Our study of the spin resolved electronic structure of ZGNRs is based on the {\it ab initio} 
pseudopotential density functional method~\cite{21} within the local spin density 
approximation~\cite{22}. 
A periodic saw-tooth-type potential perpendicular to the direction of the 
ribbon edge is used to simulate the external electric fields ($E_\text{ext}$) 
in a supercell (Fig. 1). 
In accordance with previous convention~\cite{12,13,14,15}, 
the ZGNRs are classified by the number of zigzag 
chains ($n$; Fig. 1) forming the width of the ribbon. 
We will hereafter refer to an ZGNR with $n$ zigzag 
chains an $n$-ZGNR. 
When the spin degree of freedom is neglected, 
our calculation from first-principles also 
predicts a two-fold degenerate flat band at $E_F$ (Fig. 2a). 
But the spinless state is not the ground state. 
Moreover, the electronic structures of the ZGNRs show 
marked alterations when spins and $E_\text{ext}$ are included.

\begin{figure}[b]
\centering
\includegraphics[width=8.5cm]{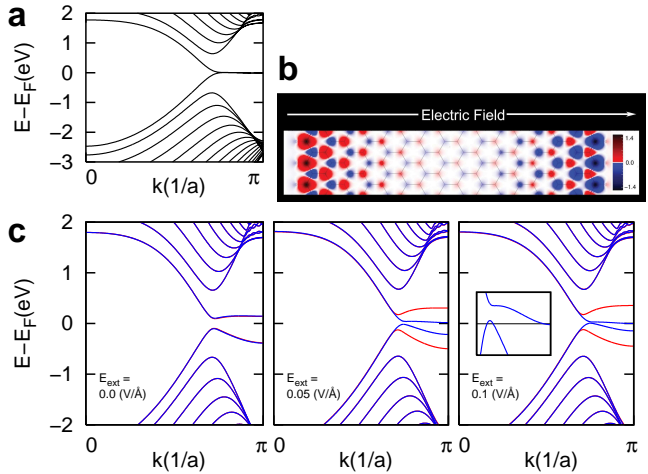}
\caption{
{\bf Electronic structures of graphene nanoribbons.} 
In all the figures, the Fermi energy ($E_F$) is set to zero. 
{\bf a}, The spin-unpolarized band structure of an 16-ZGNR. 
{\bf b}, The spatial distribution of the charge difference between $\alpha$- 
and $\beta$-spin ($\rho_\alpha(r) -\rho_\beta(r)$) 
for the ground state when there is no external field. 
The magnetization per edge atom for each spin on each sublattice 
is 0.43 $\mu_B$ with opposite orientation,
where $\mu_B$ is the Bohr magneton. 
The graph is the density integrated in $z$ direction 
and the scale bar is in unit of
$10^{-2}|e|/$\AA$^2$. 
{\bf c}, From left to right, the spin-resolved band structures of an 
16-ZGNR with $E_\text{ext}$ = 0.0, 0.05, and 0.1 V/\AA,~respectively. 
The red and blue lines denote bands of $\alpha$-spin and $\beta$-spin states,
respectively.
Inset, the band structure 
with $E_\text{ext} = 0.1$ V/\AA~in the range $|E- E_F| < 50$ meV and 
$0.7\pi \le ka \le \pi$ (the horizontal line is $E_F$). 
}
\end{figure}

Considering first the spin degree of freedom, we find as in previous studies that the 
configuration with opposite spin (antiferromagnetic) orientation between ferromagnetically 
ordered edge states at each edge (Fig. 2b) is favored as the ground state over the 
configuration with same spin orientation between the two edges~\cite{12,18,19}. 
(The present result of 
antiferromagnetic spin configuration on the honeycomb lattice is consistent with
a theorem for electrons on bipartite lattice~\cite{23}.) 
Our calculations show that the magnetic 
interaction energies are quite large. For example, the total energy difference between a 
spin-polarized edge and a spin-unpolarized one is 20 meV per edge atoms in the case of 
8-ZGNR, and the spin configuration is further stabilized by 2.0 meV per edge atom due to 
the antiferromagnetic coupling between the spin-polarized edges. 
Because the interaction 
between spins on opposite edges increases with decreasing width, the total energy of an 
$n$-ZGNR with antiferromagnetic arrangement across opposite edges is always lower than that 
of a ferromagnetic arrangement if $n\le 32$. This total energy hierarchy is maintained when 
external electric fields are applied. 
It is known that spontaneous magnetic orderings 
in one- and two-dimensional spin lattice model are difficult to
achieve at finite temperature~\cite{24}. 
Spin correlation lengths comparable to nanoscale systems,
however, is possible in practice~\cite{25,26,27,28}. 
Here, we also expect that spin orderings are realizable because of 
the large anisotropic exchange interactions between the spins
in ribbons with split-gate geometry on the substrate. 

We find that the ground state of the ZGNRs, including spin degree of freedom, has a 
bandgap which is inversely proportional to the ribbon width. However, the energy
splitting at $ka = \pi$ is $\sim$0.52 eV regardless of width if 
$n \ge 8$. The states of opposite spin orientation are 
degenerate in all bands (Fig. 2c, left).
When spins are included, the degeneracy between the occupied
and unoccupied edge-state bands at $E_F$ is now lifted
and the edge states near $E_F$ have dispersion along the direction
of the edge with a band width of $\sim$2 eV when extended over 
the Brillouin zone.

With applied transverse electric fields, we find that the valence and conduction 
edge-state bands associated with one spin orientation 
close their gap, whereas those associated with 
the other widen theirs (Fig. 2c). 
So, under appropriate field strengths, the ZGNRs are forced 
into a half-metallic state by the applied electric field, 
resulting in insulating behavior for 
one spin and metallic behavior for the other. 
We shall defer the discussion of spin-orbit interactions later. 
For now, we label the gap-opening states as $\alpha$-spin (shown in red in Figs 2-4)
and the gap-narrowing states as $\beta$-spin (blue). 
In a 16-ZGNR, the bandgap associated with $\beta$-spin is 
completely closed by an $E_\text{ext}$ = 0.1 V/\AA,~whereas the gap for 
$\alpha$-spin electrons remains very large at 0.30 eV (Fig. 2c). 
The energy gap for the $\beta$-spin electrons 
changes to an indirect gap from a direct gap as $E_\text{ext}$ increased,
and is closed indirectly (Fig. 2c, inset). 
After gap closure, an electron channel near $ka = \pi$ 
and a hole channel near $ka = 0.75\pi$ appear at $E_F$, 
all with the same spin direction. 

\begin{figure}[b]
\centering
\includegraphics[width=8.5cm]{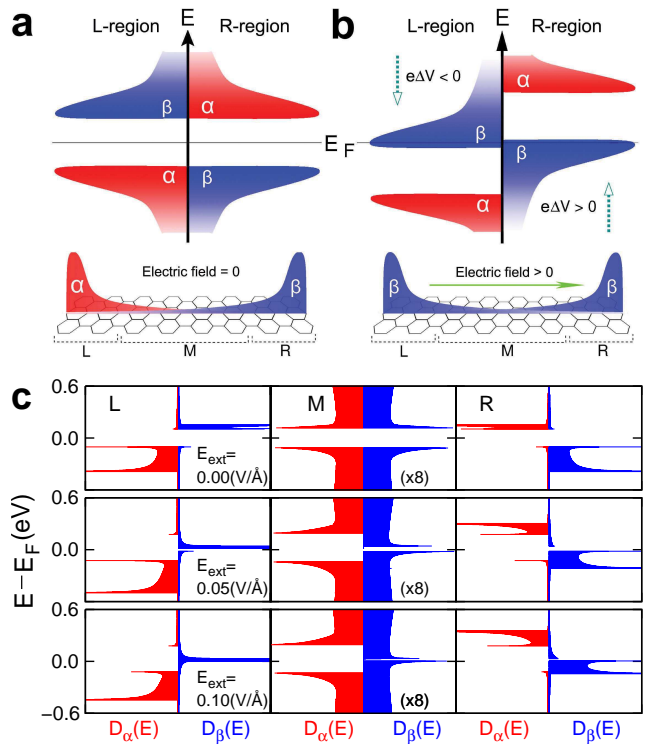}
\caption{
{\bf Origin of half-metallicity.} 
{\bf a}, Schematic density of states diagram of the 
electronic states of a ZGNR in the absence of an applied electric field. 
Top: the occupied and unoccupied localized edge states 
on left side (L-region as defined at the bottom) 
are $\alpha$-spin and $\beta$-spin states, respectively, and
and vice versa on right side (R-region) with the same energy gap for both 
sides. Bottom: schematic diagram of the spatial spin distribution 
of the highest occupied valence band states without an external 
electric field. 
{\bf b}, Top: with a transverse electric field, 
the electrostatic potential on the left side is lowered ($e\Delta V<0$) 
while the one on the right side raised ($e\Delta V>0$). 
Correspondingly, the energies of the localized state at the
left edge are decreased and those of the localized states at right edge increased.
Bottom: the resulting states at $E_F$ are only $\beta$-spin. 
{\bf c}, From left to right, the local density of states of $\alpha$- 
and $\beta$-spins (ordinate) of a 16-ZGNR as a function of energy (abscissa) 
for atoms in the L, M and R regions shown in {\bf a}, respectively. 
From top to bottom, $E_\text{ext}$= 0.0, 0.05 and 0.1 V/\AA~respectively. 
The local density of states in the middle panels are 
enlarged eightfold for clarity. For $E_\text{ext}=0.1$ V/\AA, 
the van Hove singularities of $\beta$-spin 
in the M and R region are above the
$E_F$ by 5 meV, and all states at $E_F$ are of $\beta$-spin. 
}
\end{figure}

The half-metallicity of the ZGNRs originates from the fact that the applied electric 
fields induce energy level shifts of opposite signs for the spatially separated spin ordered 
edge states. Such separate and opposite energy shifts are made possible by the localized 
nature of the edge states around $E_F$. 
Because oppositely oriented spin states are located at the 
opposite sides of the ZGNR, 
the effect of $E_\text{ext}$ on them is opposite, moving the occupied and 
unoccupied $\beta$-spin states closer in energy but moving the occupied and 
unoccupied $\alpha$-spin states apart (Fig. 3). 
The electrostatic potential is raised on the right side and lowered on the 
left side as $E_\text{ext} (>0)$ increases. 
Correspondingly, the energies for localized edge states on 
the right side are shifted upward and those on the left side downwards,
eventually leaving  
states of only one spin orientation at $E_F$ (Fig. 3b). 
In a 16-ZGNR, the occupied $\alpha$-spin states and 
unoccupied $\beta$-spin states on the left side move downwards
in energy by 19 meV and 110 meV, respectively,
and occupied $\beta$-spin and unoccupied $\alpha$-spin states on the right side 
upwards by 112 meV and 74 meV, respectively, as $E_\text{ext}$ 
increases to 0.1 V/\AA~(Fig. 3c). 
The occupied $\beta$-spin states in the middle of a 16-ZGNR are the tails of 
the localized $\beta$-spin states on the right side and the unoccupied $\beta$-spin states 
are from the left side,
so that occupied and unoccupied $\beta$-spin states in the middle of the ZGNR move 
oppositely to close the gap. 
The energies of the occupied and unoccupied $\alpha$-spin states in the middle 
also follow movements of those of the corresponding localized states on each side, 
resulting in an increased gap value (Fig. 3c). 

\begin{figure}[b]
\centering
\includegraphics[width=6.5cm]{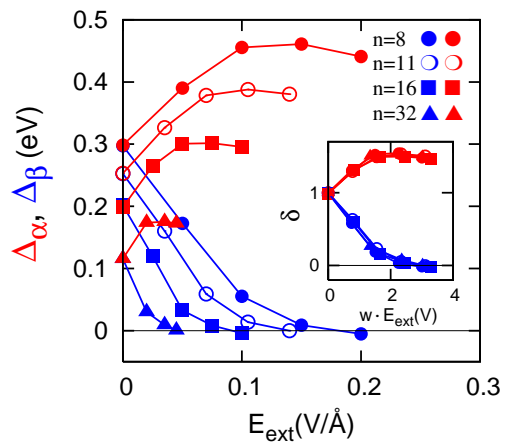}
\caption{
{\bf Dependence of half-metallicity on system size}. 
$\Delta_\alpha$ (red) denotes the direct 
band gap of $\alpha$-spin, and $\Delta_\beta$ (blue) the (in)direct gap of $\beta$-spin 
as function of $E_\text{ext}$ 
for 8-ZGNR (filled circles), 11-ZGNR (open circles), 
16-ZGNR (squares), and 32-ZGNR (triangles). 
The slope variation of $\Delta_\alpha$ and $\Delta_\beta$
indicates gap change from direct to indirect.
The rescaled gap $\delta_{\alpha}(w,E_\text{ext})
\equiv{\Delta_{\alpha}(w,E_\text{ext})}/{\Delta_0(w)}$ and 
$\delta_{\beta}(w,E_\text{ext})
\equiv{\Delta_{\beta}(w,E_\text{ext})}/{\Delta_0(w)}$
for the various widths collapse to a single function 
of $wE_\text{ext}$ as shown in the inset,
where $\Delta_{\alpha}(w,E_\text{ext})$ and 
$\Delta_{\beta}(w,E_\text{ext})$ are the 
bandgap for $\alpha$-spins and $\beta$-spins, respectively,
of the ZGNR with a width of $w$ in $E_\text{ext}(\neq 0)$, 
and 
$\Delta_0(w)\equiv\Delta_{\alpha}(w,E_\text{ext}=0)=
\Delta_{\beta}(w,E_\text{ext}=0)$.
}
\end{figure}

We note that the critical electric field for acheiving half-metallicity in ZGNRs  
decreases as the width increases because the electrostatic potential difference between 
the two edges is proportional to the system size. 
For a 32-ZGNR whose width is 67.2\AA, 
$E_\text{ext}$  = 0.045 V/\AA~is required to close the band gap 
for the $\beta$-spin electrons (Fig. 4). 
Because the energy shifts of the edge states depends on the total voltage drop between the 
two sides, the variation of the energy gap is expected to exhibit a universal behavior as 
function of $wE_\text{ext}$ 
where $w$ is the width of the ZGNR. This is seen in the inset in Fig. 4. 
From the calculations, the required critical field is estimated to be 3.0
(V)/$w$(\AA). 
To establish half-metallicity, 
the relevant energy scale is given by the field-induced energy 
shift, and its magnitude is in the order of 100 meV. 
Thus the small magnitude of spin-orbit interaction(4$\sim$6 meV) 
in carbon atoms~\cite{29,30} would not change the half-metallic nature of 
the ZGNRs, but would function in determining the spatial direction 
(normal direction with respect to the ribbon plane) of spin up and down 
in the ZGNRs~\cite{29}.

Because edges are inevitably susceptible to defects, we have examined the robustness of 
the predicted half-metallicity to edge defects. Our calculations show that the system 
remains purely of one spin-type at $E_F$ in the presence of different type and concentration of 
defects. Results on 8-ZGNRs with three different kinds of defects 
(dangling bonds, vacancies, and Stone-Wales defects at 6$\sim$12\% defect 
concentration per edge) are presented in Supplementary Fig. 1, confirming that the 
predicted half-metallicity is indeed robust.
 
Another consideration is that, 
when in the half-metallic state the ZGNRs are in a transverse electric field, the 
current-carrying electrons moving from the source to the drain in the longitudinal
direction would experience an effective magnetic field due 
to spin-orbit interactions~\cite{6,7} 
and the spins are expected to rotate. 
However, we find that the resulting extremely weak effective magnetic field 
are paralle to the spatial spin direction ($z$-direction in Fig. 1)
already determined by the instrinsic spin-orbit interaction of
carbon atoms.  
Supposed that we have $\beta$-spin electrons moving with velocity $\vec{v}=v\hat{y}$ 
in $\vec{E}=E_\text{ext}\hat{x}$, 
the effective magnetic field exerted on the $\beta$-spin electrons would be
$\vec{B}_\text{eff}=\frac{ev\hbar}{4mc^2} E_\text{ext}\hat{z}$
where $\hbar$ is the Planck constant, $m $ is the mass of an electron, 
$e$ is the charge on an electron,
and $c$ is the speed of light. 
At a critical electric field of 0.045 V/\AA~ for a 32-ZGNR, 
the estimated energy for spin-orbit coupling due to $E_\text{ext}$ is 
only $1.1\times 10^{-4}$ meV. 
We also find that, as a result of the energy gap asymmetry for each spin, 
there is no spin precession even when the direction of 
$E_\text{ext}$ is tilted or when the spatial spin direction
is altered by spin-orbit interaction arising from the substrate.
So,the spatial spin direction once determined 
would not change even if 
a strong transverse electric field is applied. 
This implies that, if we change the direction of $E_\text{ext}$, 
the spin-polarity of the carriers at $E_F$ of the 
half-metallic ribbon will be reversed because the induced 
potentials at the edges change their signs. 
Hence, under these conditions, the half-metallic nature is robust 
even though a transverse electric fields is applied, and spin polarized current should be 
obtained in transport experiment with split-gates.

\vspace{1cm}
{\bf Acknowledgements}
We thank J. Neaton, F. Giustino, I. Souza, C. H. Park 
and H. J. Choi for discussions. 
This research was supported by National Science Foundation (NSF) 
and by the Director, Office of Science, 
Office of Basic Energy Science, Division of 
Material Sciences and Engineering, 
U.S. Department of Energy (DOE). 
Computational resources have been provided 
by the NSF at the National Partnership for Advanced Computational
Infrastructure and by the DOE at the National Energy Research Scientific
Computing Center.

{\bf Author Information}
The authors declare no competing financial interests.
Correspondence and requests for materials should be 
addressed to S.G.L. (e-mail: sglouie@berkeley.edu).

\newpage
\bf{Supplementary Figure and Legend}
\begin{figure}[b]
\centering
\includegraphics[width=8.5cm]{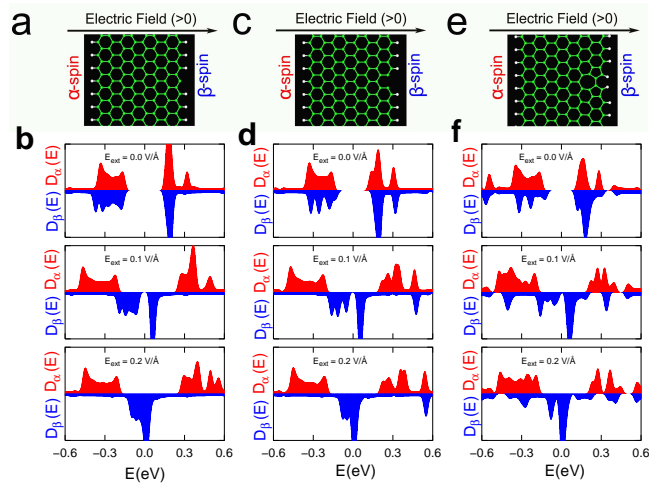}
\caption{
{\bf Supplementary Fig1. Robustness of half-metallicity in defective graphene 
nanoribbons.} {\bf a}, 
A ball-and-stick model of an 8-ZGNR with one dangling bond (an 
edge carbon atom without hydrogen passivation) on the right edge. One dangling 
bond per 17 edge atoms are considered, which corresponds to 5.9\% impurity 
concentration per one edge. In the figure, the atomic structure near the defect is 
displayed. Electric fields ($E_\text{ext}>0$) 
are applied from the left side to the right side. The 
$\alpha$-spin state is located on the left side and the $\beta$-spin state 
on the right side in the case of the ground state without 
applied electric fields. {\bf b}, From top to bottom panels, the spin 
resolved total density of states (TDOS) are drawn for a defective 8-ZGNR shown in 
{\bf a} with $E_\text{ext}$ = 0.0, 0.1, and 0.2 V/\AA~respectively. 
At the same critical $E_\text{ext}$ of 0.2 V/\AA~for an ideal 8-ZGNR, 
the gap for $\beta$-spin state is completely closed. {\bf c}, A ball-and-stick 
model of an 8-ZGNR with one carbon-atom vacancy on the right edge. 
Defect concentration, ground spin configuration, and 
direction of electric fields are identical to the case of 
an 8-ZGNR with one dangling bond shown in {\bf a}. {\bf d}, The TDOS for an 8-ZGNR with 
one vacancy on the right edge (shown in {\bf c}) with and without electric fields. 
Half-metallic nature persists also in this case. {\bf e}. A ball-and-stick model for an 
8-ZGNR with one single rotated bond (Stone-Wales defect at the right edge). 
Impurity concentration per one edge in this case is 11.8\% 
since two edge atoms participate to create a 
Stone-Wales (SW) defect. Electric fields and ground spin 
configuration follow the same convention described in {\bf a} and {\bf c}. 
The total energy of an 
8-ZGNR with a SW defect is much higher than that of the ideal one by 
3.05 eV per defect so that appropriate treatments (e.g. annealing) on the 
sample will remove this highly unstable defect of this kind. 
Nevertheless, even with such high impurity concentration shown in {\bf e}, 
the TDOSs displayed in {\bf f} clearly show the 
robustness of the half-metallicity since the gap for the $\beta$-spin states 
is completely closed while the gap for $\alpha$-spin state 
is at 0.39 eV with $E_\text{ext}$ = 0.2 V/\AA.}
\end{figure}

\end{document}